# CG-3DSRGAN: A classification guided 3D generative adversarial network for image quality recovery from low-dose PET images

Yuxin Xue, Yige Peng, Lei Bi, and Dagan Feng, *Fellow, IEEE,* Jinman Kim, *Member, IEEE*

*Abstract*— Positron emission tomography (PET) is the most sensitive molecular imaging modality routinely applied in modern healthcare. High radioactivity caused by the injected tracer dose is a major concern in PET imaging and limits its clinical applications. However, reducing the dose leads to inadequate image quality for diagnostic practice. Motivated by the need to produce high quality images with minimum 'low-dose', convolutional neural networks (CNNs) based methods have been developed for high quality PET synthesis from its low-dose counterparts. Previous CNNs-based studies usually directly map low-dose PET into features space without consideration of different dose reduction levels. In this study, a novel approach named CG-3DSRGAN (Classification-Guided Generative Adversarial Network with Super Resolution Refinement) is presented. Specifically, the use of a multi-tasking coarse generator, guided by a classification head, allows for a more comprehensive understanding of the noise-level features present in the low-dose data, resulting in improved image synthesis. Moreover, to recover spatial details of standard PET, an auxiliary super resolution network – Contextual-Net – is proposed as a second-stage training to narrow the gap between coarse prediction and standard PET. We compared our method to the state-of-the-art methods on whole-body PET with different dose reduction factors (DRF). Experiments demonstrate our method can outperform others on all DRF.

*Clinical Relevance*— Low-Dose PET, PET recovery, GAN, task driven image synthesis, super resolution

## I. INTRODUCTION

Positron Emission Tomography (PET), an ultrasensitive and non-invasive molecular imaging technique, has been the main instrument for oncology [1] and neurology [2]. Compared with other imaging modalities, such as Magnetic Resonance (MR) and Computed Tomography (CT), PET can image the functional properties of living tissue and detect disease-related functional activity within organs by injecting radioactive tracers into the body [3]. Unfortunately, the ionizing radiation dose from the injected radioactive tracer to the patient that is necessary for PET imaging, limits its application [4]. According to the dose level of the tracer, the reconstructed PET images can be classified as standard-dose (sPET) and low-dose PET (lPET) images. When compared to the lPET counterpart, the sPET images have higher signal-to-noise ratios (SNR), more structural features, and superior image quality overall. However, sPET will unavoidably have high cumulative radiation exposure. Motivated by these challenges, there have been developments in image analysis methods that aim to recover sPET images from lPET images while maintaining a low injected dose.

Y. Xue, Y. Peng, L. Bi, D. Feng, and J. Kim are with the School of Computer Science, the University of Sydney, Australia. (Corresponding: jinman.kim@sydney.edu.au).

Deep learning methods based on convolutional neural networks (CNNs) have achieved great success in medical image analysis related tasks e.g., automated tumor segmentation and classification [5–6]. Investigators have also used deep learning for recovering sPET from lPET images [7-9]. Xiang et al. [8] integrated multiple CNN modules following an auto-context strategy to estimate sPET from lPET via multiple iterations; they adopted the U-Net architecture dilated kernels to increase the receptive field. Unfortunately, the pooling layers in CNNs are used to reduce the spatial resolution of the feature maps, which can cause information loss and fine-grained details e.g., edge and texture. To address these challenges, investigators have attempted to use generative adversarial network (GAN) to preserve structural details by extending the network with a discriminator to distinguish true/synthetic images [9]. Bi et al. [10] developed a multi-channel GAN to synthesize high quality PET from corresponding CT scan and tumor label. Using a conditional GAN model and an adversarial training strategy, Wang et al. were able to recover full-dose PET images from low-dose PET [11]. However, GAN-based approaches have difficulties in recovering high dimensional details such as contextual information and clinically significant texture features (e.g., image intensity values). This is mainly attributed to the fact that these methods have not explored the spatial correlations between the sPET and lPET images. To overcome the artifacts caused by GAN, a second-stage refinement is an ideal way to rectify the mapping error caused in the previous stage. Inspired by this unmet need, we delicately designed a super resolution network – Contextual-Net to reconstruct more contextual details so that the refined synthetic PET is spatially aligned with real sPET.

Task driven strategy has been actively investigated in medical image analysis field [12-14]. Basically, task-driven approaches introduce additional supervision which is semantic task-related to boost main model performance. Zhang et al. [12] proposed a TD-GAN which aims to convert 3D CT scans into X-rays with a segmentation-driven strategy. Chaitanya et al. [13] reversely used a synthesis-driven data augmentation strategy to improve cardiac MRI segmentation. Tang et al. [14] adopted a task-oriented method by leveraging an image translation framework to improve unsupervised domain adaptation results on paediatric pneumonia recognition. The above task-driven approaches have demonstrated benefits for the primary task that are semantically related. In our work, since lPET has the property of various dose reduction factors, we employ a classification guided strategy to enhance the performance of sPET synthesis.

In this study, we proposed a 3D classification guided super resolution generative adversarial network (CG-3DSRGAN) to recover sPET images from corresponding lPET images. We adopted a task-driven strategy to guide the main PET image synthesis task by categorizing the input lPET into certain dose reduction levels while the reconstruction process. Additional super resolution component – Contextual-Net was embedded with coarse generator and discriminator in CG-3DSRGAN to further recover contextual details of synthetic sPET. Our main contributions are as follows:

• An end-to-end generalized GAN-based super resolution framework, CG-3DSRGAN, is proposed to synthesize sPET quality level from different dose reduced lPET.

• A task-driven classification sub-branch is ensembled to the coarse generator to form a multi-tasking network termed ML-Net. By training the main synthesis task and proxy classification task in parallel, ML-Net simultaneously learns generalized features and dose reduction specific features.

• CG-3DSRGAN adopts a coarse-to-fine synthesis strategy by incorporating a coarse generator ML-Net with super resolution Contextual-Net. This progressive method can preserve more high-dimensional details and obstacle over-smooth issue caused by GAN.

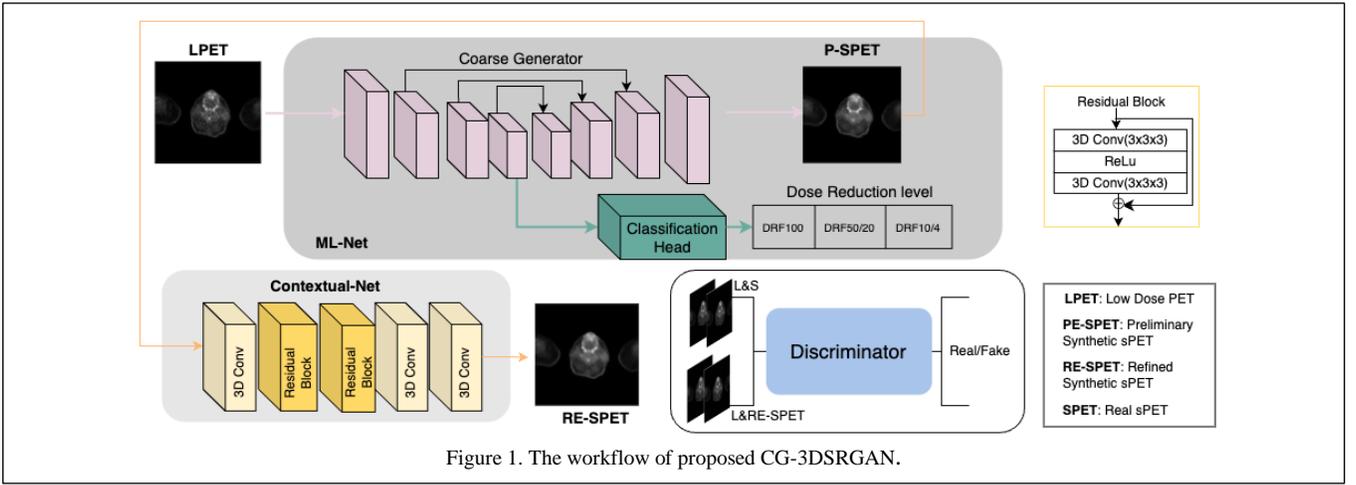

Figure 1. The workflow of proposed CG-3DSRGAN.

## II. METHODS

The framework of our CG-3DSRGAN is shown in Figure 1 consists of three components: a) a multi-task reconstruction network – ML-Net to generate an initial prediction result that closely resembles the actual sPET image and classification outputs that predicts input image dose reduction level; b) a super resolution network – Contextual-Net to refine the first stage result with the objective of preserving high-dimensional features and contextual details; and c) a discriminator to distinguish the veracity of the refined PET.

### A. Task-driven coarse synthesis model – ML-Net

ML-Net consists of two parts: the main reconstruction network $R$ which adopts a 3D-UNet [15] like structure and a proxy classification head which share the same encoder. $R$ synthesizes sPET $\tilde{y}_s$ from corresponding lPET $x$ and classification head projects input into three-dimensional vector to predict lPET dose reduction level $\tilde{y}_c$.

The encoder of ML-Net is composed of five down-sampling blocks, and each of them adopts a 3 x 3 x 3 convolutional kernel with stride 2. The encoder blocks are introduced in the form of Leak ReLU – Convolutional layer – Batch Normalization (BN). The decoder structure also contains five up-sampling blocks with convolutional kernel 3 x 3 x3 and stride 2. The decoder blocks are made up of three sequential parts: ReLU, a transposed convolutional layer, and BN. The classification head is made of multi-layer perceptron.

Two supervisions are used for the ML-Net training, one is to enforce the coarse synthetic sPET $\tilde{y}_s$ to be mapped as real sPET $y_s$ at the pixel level. The reconstruction objective is defined by L1 loss which can be formulated as follow:

$$L_{re} = \|\tilde{y}_s - y_s\|_1. \quad (1)$$

Another constraint is added to classification branch. Given the true category of input lPET reduction level $y_c$, a cross-entropy loss is used for the dose reduction level prediction task:

$$L_{class} = -y_c \log \tilde{y}_c. \quad (2)$$

### B. Refine Network-Contextual-Net

Though the synthetic sPET $\tilde{y}_s$ from ML-Net is similar to real sPET in structure and content, there still is a mapping gap between them. To obtain clinical-level estimated sPET, we employ a refine network-Contextual-Net to estimate the lost contextual information between $\tilde{y}_s$ and real sPET $y_s$ in a super-resolution manner.

As shown in Fig.1, Contextual-Net consists of a convolutional layer (Conv) with a kernel size of 3 x 3 x3 and stride 1 and residual blocks (RB) in the form of Conv-RB-RB-Conv-Conv. The residual block is constructed by the Conv-ReLu-Conv structure and the final output is incorporated with the input.

Refined synthetic sPET $\tilde{y}_s'$ is encouraged to be super resolution output. Thus, we adopt perceptual loss [16] to further narrow the gap between $\tilde{y}_s'$ and ground truth $y_s$ in textural and contextual level:

$$L_{refine} = \sum_i^I \frac{1}{C_i H_i W_i} \|\varphi_i(\tilde{y}_s') - \varphi_i(y_s)\|_2^2 \quad (3)$$

Where $\varphi_i$ is the activation of $i$th layer of pre-trained 16-layer VGG $\varphi$. $I$ denotes a set of activation layers.

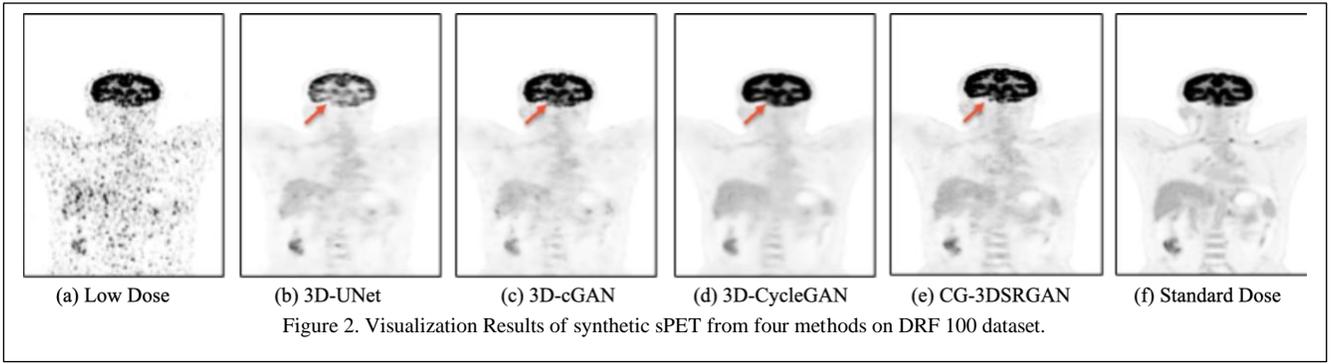
Figure 2. Visualization Results of synthetic sPET from four methods on DRF 100 dataset.

## C. Discriminator

The Discriminator $D$ receives paired images, including the lPET $x$ and the corresponding real/ refined synthesized sPET to distinguish the fake pair from the real one. The discriminator has a similar structure to that of pix2pix [20] while the 2D convolutional layer is replaced by 3D operations. Adversarial loss is used to push fake/real pairs to be undistinguishable for discriminator $D$:

$$L_{dis} = E_{l,s}[(D(x, y_s) - 1)^2] + E_l\left[\left(D(x, \tilde{y}'_s)\right)^2\right] \quad (4)$$

Based on all the above, the overall loss function for CG-3DSRGAN is formulated as below:

$$L_{total} = \lambda_1 L_{re} + \lambda_2 L_{class} + \lambda_3 L_{refine} + \lambda_4 L_{dis} \quad (5)$$

## III. EXPERIMENTAL RESULTS

### A. Dataset Description

We used dataset from the Ultra-Low Dose challenge [21]. It consists of 281 studies of whole-body PET images, which were acquired from the United Imaging uEXPLORER scanner. All data were acquired in list-mode allowing for rebinding of data to simulate different acquisition times. Simulated low-dose PET images have a specific dose reduction factor (DRF), which represents the reduced counts in the resampled time window used during the reconstruction process. Higher DRF means more reduced acquisition time, the corresponding lPET has more defects. Low-dose images will be provided with DRF at 4, 10, 20, 50 and 100, as well as standard-dose images.

### B. Experimental Settings

The CG-3DSRGAN was trained using a batch size of 4 with the Adam optimizer. We empirically set $\lambda_1 = 300$, $\lambda_1 = 10$, $\lambda_1 = 10$ and $\lambda_4 = 1$ for the hyper-parameter defined (5) and were fixed in the subsequent tests. We trained the proposed method for 100 epochs The learning rate was initially set as 2e-4, which was then linearly decreased with a factor of 0.1 and patience of 5 epochs. To avoid overfitting, an early stopping strategy was applied and was used to terminate the training process when the learning rate exceeds 2e-6. All the experiments were conducted on a 12GB NVIDIA GeForce RTX 2080Ti GPU, with the PyTorch framework.

### C. Comparison with the state-of-the-art methods

We compared our method with the state-of-the-art PET synthesis methods including 3D-UNet, 3D-cGAN [11] and 3D-CycleGAN [19] which also had good performance in Ultra-low Dose challenge [18]. As shown in Table I, the CG-3DSRGAN outperformed the other three methods with overall better reconstruction results on all DRF datasets. ∂When compared to 3D-UNet, our proposed CG-3DSRGAN achieved substantial improvements over all evaluation metrics, specifically, the PSNR was improved from 44.770dB to 48.453dB on DRF 100 while the NRMSE decreased from 1.402% to 0.342%. As for the other two GAN-based models (3D-cGAN and 3D-CycleGAN), CG-3DSRGAN demonstrated its advantage of generating high quality synthetic PET with an increase of 4.134dB and 3.537dB in PSNR on DRF 4, and an improvement of 3.855dB and 2.389dB in PSNR on DRF 10, respectively. It is worth mentioning that compared with the second-optimal method, 3D-CycleGAN, our proposed method presented a superior performance on pixel-level reconstruction results by reducing the NRMSE from 0.495% to 0.342% on DRF 100 and optimizing NRMSE from 0.398% to 0.265% on DRF 50.

Qualitative results on DRF 100 dataset of the comparison methods are shown in Figure 2 which demonstrates that CG-3DSRGAN has superiority in reconstructing structure and recovering spatial information, especially in the brain structure and liver texture.

TABLE I. QUANTITATIVE COMPARISON RESULTS OF THE IMAGES SYNTHESIZED BY DIFFERENT METHODS FOR DIFFERENT DRFs

|  | *Methods* | *PSNR* | *SSIM* | *NRMSE* |
|---|---|---|---|---|
| DRF4 | Low-Dose | 58.718 | 0.999 | 0.154 |
|  | 3D-UNet | 52.871 | 0.995 | 0.317 |
|  | 3D-cGAN | 54.487 | 0.996 | 0.284 |
|  | 3D-CycleGAN | 55.084 | 0.997 | 0.240 |
|  | **CG-3DSRGAN** | **58.621** | **0.998** | **0.161** |
| DRF10 | Low-Dose | 54.421 | 0.997 | 0.249 |
|  | 3D-UNet | 49.481 | 0.993 | 0..424 |
|  | 3D-cGAN | 52.315 | 0.994 | 0.319 |
|  | 3D-CycleGAN | 53.781 | 0.994 | 0.343 |
|  | **CG-3DSRGAN** | **56.170** | **0.999** | **0.174** |
| DRF20 | Low-Dose | 51.698 | 0.995 | 0.347 |
|  | 3D-UNet | 49.231 | 0.994 | 0.436 |
|  | 3D-cGAN | 51.827 | 0.995 | 0.343 |
|  | 3D-CycleGAN | 52.418 | 0.995 | 0.307 |
|  | **CG-3DSRGAN** | **54.478** | **0.998** | **0.206** |
| DRF50 | Low-Dose | 47.490 | 0.983 | 0.597 |
|  | 3D-UNet | 48.149 | 0.993 | 0.501 |
|  | 3D-cGAN | 48.745 | 0.994 | 0.426 |
|  | 3D-CycleGAN | 49.527 | 0.995 | 0.398 |
|  | **CG-3DSRGAN** | **51.867** | **0.996** | **0.265** |
| DRF 100 | Low-Dose | 42.490 | 0.970 | 1.402 |
|  | 3D-UNet | 44.770 | 0.992 | 0.744 |
|  | 3D-cGAN | 45.904 | 0.993 | 0.625 |
|  | 3D-CycleGAN | 47.282 | 0.993 | 0.495 |
|  | **CG-3DSRGAN** | **48.453** | **0.994** | **0.388** |

*D. Ablation Study*

To evaluate the effectiveness of important components in our proposed CG-3DSRGAN, we conducted the following ablation study, including a) baseline, reconstruction results of the coarse generator; b) baseline with classification guided strategy (baseline + CG); c) baseline + CG + Discriminator (CG-3DGAN); d) the proposed CG-3DSRGAN. The most challenging lPET images with DRF 100 were used for the ablation study.

To assess the performance of our task-driven strategy, we compared the results between the baseline with/without CG. As shown in Table II, Classification guided strategy improved the baseline by 0.813, 0.002, 0.445% in PSNR, SSIM and NRMSE respectively. It indicated that the proposed task-driven strategy was effective for high quality PET synthesis on the structure and spatial information. We argued that the classification head connected with the encoder of the coarse generator could enhance the feature representation for lPET in different dose reduction levels, thus further boosting synthesis model generalization capability.

By adding a discriminator into the baseline + CG, PSNR improved to 46.832dB from 46.019dB and NRMSE dropped to 0.430%. GAN structure showed the advantage of structure level recovery by increasing SSIM to 0.994.

To investigate how Contextual-Net contributed to sPET synthesis results, it was added to CG-3DGAN, resulting in a large improvement of 0.982dB in PSNR and 0.018% NSMSE correction.

TABLE II. QUANTITATIVE COMPARISON RESULTS OF DIFFERENT VARIANTS OF CG-3DSRGAN

| Methods | DRF100 | | |
|---|---|---|---|
| | **PSNR** | **SSIM** | **NRMSE** |
| Baseline | 46.019 | 0.990 | 0.475 |
| Baseline+CG | 46.832 | 0.992 | 0.430 |
| CG-3DGAN | 47.471 | 0.994 | 0.406 |
| CG-3DSRGAN | 48.453 | 0.994 | 0.388 |

IV. CONCLUSION

In this work, we proposed a novel high quality PET synthesis model – Classification-guided 3D super resolution GAN. Our results demonstrated the proposed method has optimal performance on standard PET synthesis from lPET quantitively and qualitatively compared to other existing state-of-the-art. As future work, we will investigate the use of residual estimation that can further improve synthesis sPET quality by minimizing the distribution gap between prediction and ground truth. A self-supervised pre-training strategy may also further improve the model representation.